\documentclass[12pt, a4paper]{amsart}

\usepackage{hyperref}
\usepackage{amsmath, amsfonts, amsthm,bm,mathrsfs}
\usepackage{amsmath}
\usepackage{natbib}
\usepackage{graphicx}
\usepackage{amscd,amssymb,amsthm}
\makeatletter
\g@addto@macro{\endabstract}{\@setabstract}
\newcommand{\authorfootnotes}{\renewcommand\thefootnote{\@fnsymbol\c@footnote}}%
\makeatother

\begin{document}
\bibliographystyle{plainnat}

\begin{center}
  \LARGE 
Trend extraction  in functional data   of  R and T waves amplitudes of  exercise electrocardiogram
\vskip20pt

 \normalsize
  \authorfootnotes
  Camillo Cammarota\footnote{cammar@mat.uniroma1.it, corresponding author}\textsuperscript{1}, Mario Curione\footnote{Mario.Curione@uniroma1.it}\textsuperscript{2} 

  \textsuperscript{1}Department of Mathematics  University La Sapienza,\\
 P.le A. Moro 5, 00185 Rome, Italy \par
  \textsuperscript{2}Internal Medicine and Medical Specialties Department,\\   
  University La Sapienza

\par \bigskip

  \today
\end{center}

\begin{abstract}
The   R and T waves amplitudes of the electrocardiogram recorded  during the exercise test undergo strong modifications in response to stress.  We analyze  the time series  of these amplitudes in  a group of normal subjects  in the framework of functional data,   performing reduction of dimensionality, smoothing and  principal component analysis. These methods show that  the  R and  T  amplitudes  have opposite responses to stress,  consisting  respectively  in  a bump and a dip  at the early recovery stage. We test these features  computing a confidence band for the  trend of the population  mean and analyzing the zero crossing of its  derivative. 
 Our findings  support the   existence of  a  relationship between R and T wave amplitudes and respectively  diastolic and systolic ventricular volumes.
\end{abstract}
 
 Keywords:
ECG;  R wave; T wave;   exercise test; time series;   trend; functional data;    extrema; maxima; minima.

\section{Introduction}
The    electrocardiogram (ECG)  recorded during  the  exercise  test is used    in clinical practice  to evaluate the presence  of myocardial ischaemia \cite{gibbons}. During the test the  patient on a bicycle ergometer is subjected to a  workload    increasing  in time (stress phase). When the heart rate reaches its maximum  (acme), the exercise is stopped and gradually the heart rate recovers its basic value (recovery phase).  The  R and T waves   of the ECG  (fig. 1) occur respectively at the end of the ventricular  diastolic phase (telediastole) and at  the end of the systolic phase (telesystole), which are  the times of maximum and minimum ventricular filling.   

Great attention was devoted in clinics  to the modifications of the QRS complex (fig. 1) during exercise. 
The QRS  amplitude    was found  to show opposite directional changes  in different studies \cite{kligfield}. These non univocal  findings are due mainly  
to two factors:  the great  inter individual variability of the response to exercise in normal subjects and the large fluctuations  in amplitude of the ECG waves that occur at the time scale of few beats.  Observations localized in time as the ones typical of  clinical research  revealed  not sufficient to extract   significant directional changes during the exercise.  
ECG and vectorcardiographic parameters  related to R and T waves  during exercise have been recently investigated
\cite{lipponen, kania, bortolan}.
 In different conditions  concomitant variations of  R and  T waves     were  observed    and  their relationship to the changes in ventricular cavity size  were conjectured  \cite{feldman}.  These variations and their relationships to changes in ventricular volume during exercise  have been not sufficiently investigated, to the best of our knowledge.

The  RR interval,  defined as the  time  interval  between  two consecutive R peaks in the ECG, is inversely related  to the heart rate.  The sequence  of RR intervals during exercise   is characterized by a V-shaped profile,  in which the minimum corresponds to the acme (see the  RR series of a group of normal subjects  in fig. 2, top).  The series of  RR and of RT intervals, that   is characterized by the same profile,   can be investigated using the standard model of decomposition of a series into trend plus noise using non parametric methods \cite{wasserman, ruppert}. 
 The trend extraction  allows to reveal directional changes in time or to  detect maxima and minima in RR and RT series  \cite{cammarota2008, cammarota2011-1, cammarota2012}. A related  approach  aimed to assess the significance of  extrema  in  one series of observations   is  the  \verb#R#  package \verb#SiZer#  (Significance of Zero crossings of  derivatives)  \cite{sizer, chauduri}.
 
 These methods  however are restricted to the analysis of individual series. Modern datasets  of medical data  are  often in the form of  longitudinal data:  one or more indices are measured repeatedly in time in a group of subjects and  one can  assume independence among  individuals  \cite{verbeke}.  When the time  resolution is sufficiently high  so that the data  reveal an underlying curve,  this type of  dataset is known as \lq functional data\rq , and it  has  recently  obtained much attention  in  statistical literature \cite{ramsay, ferraty}.  In medical datasets the population mean  of functional observations reflects  a collective behavior  of an observable as a function of time.
A typical  problem   is  the trend extraction   of the population mean aimed  to detect the  relevant  features  and to test their significance. Theoretical investigation has been devoted  to assess if  the  population mean  is non constant in time,  and to provide a  related test of significance.  This test is   based on the estimate of  a confidence band of the population mean  \cite{cao, degras,song.q, bunea, azais}. Several software tools have been introduced  to analyze functional data:  one of these is   the  \verb#R# \cite{R}   package  \verb#fda#  \cite{fda}.  Functional data methods   revealed  recently useful   in the analysis of the ECG  \cite{ieva.smmr}.

In  the present paper  we analyze the  functional data   of RR intervals and  of  R and T waves amplitudes  measured  during the exercise test of a group of normal subjects (fig. 2). The R wave amplitude is strongly linked to  
the area of QRS complex  that   was  previously investigated  for both normal and ischaemic subjects during exercise  \cite{curione2008}. It was observed  that the population mean time profile   during the early recovery phase,  i.e.   immediately  after the acme, shows  a reduction in QRS  area  values. 
 Our aim in the present work   is twofold:  1) to test  the   significance of the minimum  in  R amplitude series at  early recovery phase, previously  observed \cite{curione2008};  2) to investigate the  concomitant presence of a maximum  in  T amplitude series  and  to assess its significance.

\begin{figure}
\centering
\includegraphics[width=0.9\textwidth]{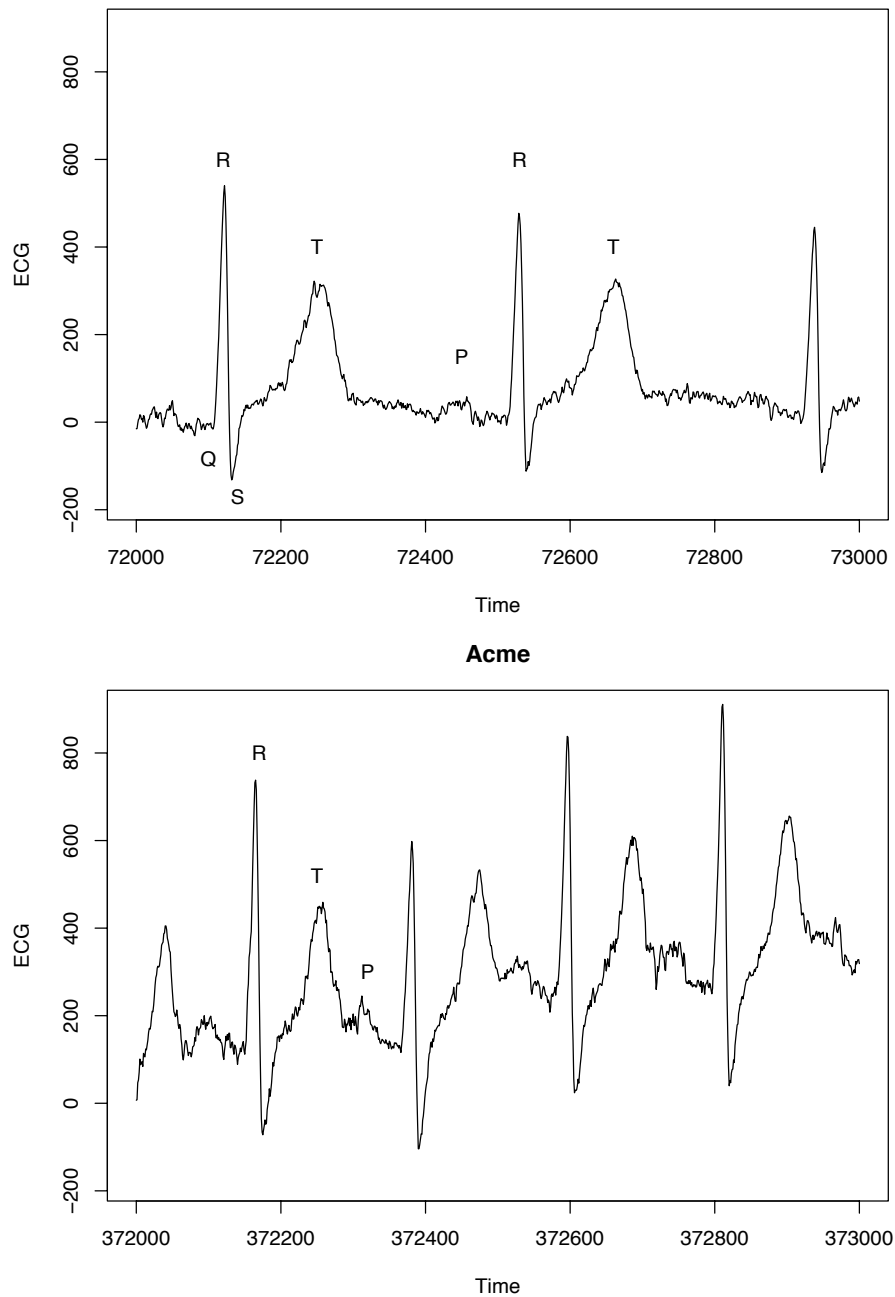}
\caption{Two-seconds  recording of  ECG   during exercise test (raw data  from the lead V5).  Top: The ECG signal  at the start of the test (rest condition) with the R peak and the  apex of T wave. Bottom: the signal at the acme, when the RR interval takes its minimum; T wave  offset overlaps with the subsequent P wave onset. Time unit = 0.002 sec; ECG  voltage resolution unit= 2.441 $\mu$V.}
\label{fig.ecg}
\end{figure}

\section{Model and methods}

\subsection{Electrocardiogram measurement and analysis}

 In multistage Bruce protocol  \cite{gibbons} the patient on a bicycle ergometer is subjected  to a  workload    increasing  in time by steps  (25 W every 2 minutes). The exercise is stopped  when the  heart rate reaches a maximum, 
usually 85\% of the estimated top heart rate based on the patient's age.  After achieving peak 
workload, the patient spends  some minutes at rest  on the bicycle until its heart rate recovers its basic value. 
 The  standard 12-leads ECG was recorded using the electrocardiograph  PC-ECG 1200 (Norav Medical Ltd.),  
which provides in output  a digital signal with resolution of 2.441$\mu$V and 500 Hz sampling frequency. The duration of the test was about ten minutes both for stress and recovery. 

 For the RR extraction the precordial lead V5 was chosen, because it is less influenced by motion artifacts (fig.1).  The R peak detection was performed using a derivative-threshold algorithm and   the onset and  offset of the QRS complex were identified. For each QRS complex  the R amplitude was defined as the maximum minus the minimum of the signal  in the window between the onset and the offset of the complex. The  local value of the baseline was defined as the mean of the two values of the signal in the onset and in the offset of the QRS. The T amplitude   was detected as the maximum of the signal   computed in a window subsequent to each R peak between 15\% and 50\% of the preceding RR interval    and  referred to the local  baseline; no negative T waves were present in our recordings.
Abnormal or undetected  beats were   less than   1\% of the total beats for each subject.  RR intervals falling outside the normal range due to undetected beats were  replaced  with the median computed over  blocks of  30 adjacent beats.  The  algorithm discarded the  R and T amplitudes related to undetected QRS complex.
Analysis of raw data, R  and T peak detection and subsequent computations were performed using the free statistical software \verb#R# \cite{R}.

\begin{figure}[!th]
\begin{center}
 \vspace*{8pt}
 \includegraphics[width=0.9\textwidth]{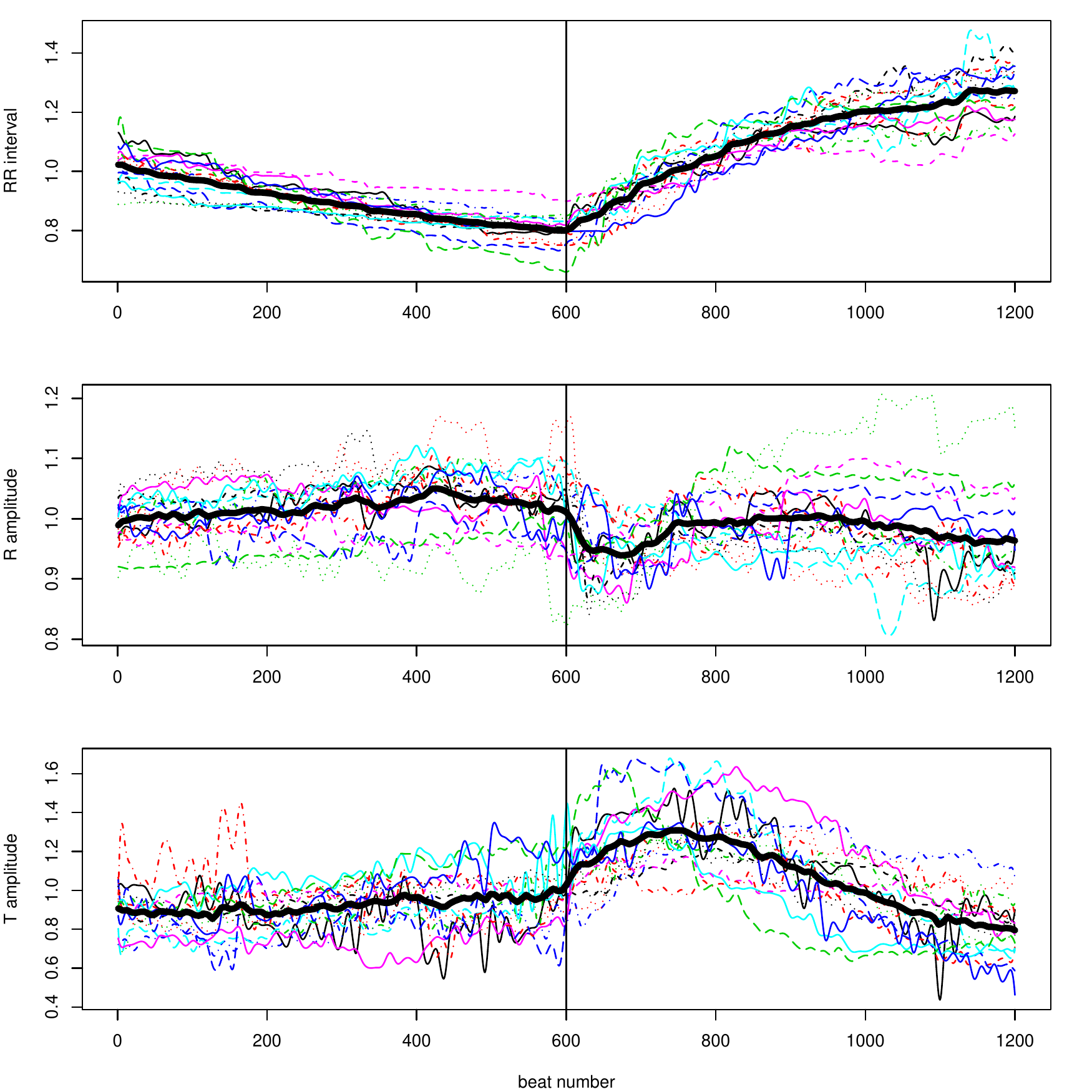}
 \end{center}
\caption{Time-mean normalized  and  aligned   series  of a group of 16 normal subjects   and their   population  mean (thick black line) of   RR interval (top), R  amplitude (centre)  and T amplitude (bottom)   during the exercise test. The vertical line denotes the acme. The series are restricted to a window of 1200 beats centered at the acme.  Adimensional units on vertical axis. Color on line.}
\label{fig.data}
\end{figure}

\subsection {Data registration and normalization}

We  denote the   data series  of the  $i$ th subject as 

\begin{equation}
X_{i} (t) , \ \ \ i=1,..., n; \ \ \ \ t=1,..., m_i
\label{rawdata}
\end{equation}
where  $t$ is the beat number index and $n$ is the number of individuals. The lengths  $m_i$ range from 1500 to 3000 beats. During the test  the duration of the RR interval, inversely related to  the 
instantaneous heart rate,  shows strong modifications: at maximal heart rate (acme) the RR interval
is about one half than the basal value. 
 Changes of amplitude of the R and T waves  occur prevalently  after the acme, but with variable delays (phase variation).
Data registration is an operation required in order to align  prominent features. We use the landmark registration based on the fact that  each RR series has a global minimum, occurring  at beat number, say  $t_i^*$, the acme.  Data registration  of R and T series is performed  putting  the time of this common  feature into a common value. This is accomplished 
extracting   a window  of  $m = 1200$  beats centered at $t_i^*$ in  each series, in such a way  that the  acme occurs at  beat number 600.  In the sequel we restrict the analysis to  this window. Fig. 2 provides a representation of the data in this window.
We assume as a model for the data  

\begin{equation}
X_i(t) = U_i\ (\mu(t) + Z_i (t) )
\label{modello}
\end{equation}
where the  population mean $\mu(t)$ is a deterministic function  of time and   $Z_i(t)$ are independent  r.v.   with zero mean and  ${\mathbb Var} Z_i(t) =\sigma_i(t)^2$, representing error measurement and  random individual deviations 
 from the population mean; the factor $U_i$ accounts for the individual amplitude of the ECG signal, that affects proportionally  the amplitudes of the R and T waves. This factor plays the same  role of an  individual  additive effect in longitudinal data models.  Since we are interested in relative variation of the variables during time,  we normalize each  series  dividing by its temporal mean, and define the normalized data as 
 \begin{equation}
Y_i(t)= \frac{X_i(t)}{\frac1m \sum_{t=1}^m X_i(t)}
\label{modellonormal}
\end{equation}
 Denoting
 \begin{equation}
 D_i = \frac1m \sum_{t=1}^m X_i(t)
 \label{effect}
 \end{equation}
and 
 $$\overline \mu = \frac1m\sum_{t=1}^m \mu(t)$$
 one has 
 $${\mathbb E}\  D_i  =  U_i\  \overline \mu\  ;  \quad {\mathbb Var}\   D_i = \frac1m U_i^2 \sigma^2_i$$
 where 
 $  \sigma^2_i= \frac 1m \sum_t \sigma_i(t)^2$. 
Since in our dataset $m$ is large    the coefficient of variation of $D_i$  
$$\frac{\sigma_i}{\sqrt m \ \overline \mu}$$  
is small, 
so  one can  approximate the r.v.   $D_i$  with its expected value   $U_i \overline \mu$.  
Hence  the normalized data can be  modeled as 
  \begin{equation}
Y_i(t)=  (\mu(t) + Z_i(t))/\overline \mu
\label{modellonormalapprox}
\end{equation}
In the normalized data the  population mean  $\mu(t)/\overline \mu)$  has a  time mean equal to 1, as shown in the three panels of fig. 2.  In the model rewritten as 
  \begin{equation}
Y_i(t)=  \tilde\mu(t) + \tilde Z_i(t)
\label{modellonormalapprox1}
\end{equation}
our aim is to detect  maxima and minima of  $\tilde\mu(t)$.

\subsection{Reduction of dimensionality and smoothing}  The basic operation for functional data is the reduction of dimensionality, that in our case is  given by $m=1200$  observations. This can be done   choosing an orthonormal set that spans a suitable functions subspace and projecting on this subspace.  We use   B-splines  made with cubic polynomials \cite{ruppert}
 $$\psi_k(t), \ \ k=1,...,K$$
  with $K$ of the order of 100.  Since these are smooth functions,  the reduction of dimensionality  produces a   smoothing of the series.
Each function $Y_i(t)$ can be  replaced  by its smoothed version $\ S_i (t)$, according to
\begin{equation}
S_i (t) = \sum_{k=1} ^K\  c_{i,k}\  \psi_k(t)
\end{equation}
where the $c_{i,k}$ are suitable coefficients. The  smoothed population  mean  $\overline{S}(t)$ is defined as 
 \begin{equation}
 \overline{S} (t)= \frac1{n} \sum_{i=1}^n  \ S_i (t) 
 \end{equation}

\subsection{Principal Component Analysis}
 The  Principal Component Analysis (PCA) in functional datasets is a method for  reduction of  dimensionality and analysis of variance, that uses a basis of principal components, the eigenfunctions of the covariance matrix. 
 Obviously the smoothed version of the data is used.  Using  the first two components   $\ \ \ \phi_1(t), \ \phi_2(t)$ 
 of this basis, denoted PC1 and PC2, 
  for each individual labeled by $i$ one has the approximation
 \begin{equation}
S_i (t) \simeq \overline{S} (t) + c_{i,1} \phi_1(t) + c_{i,2} \phi_2(t)
\label{pca}
\end{equation}
The  components PC1  and PC2  of the group and the individual  loadings $( c_{i,1}, c_{i,2})$  are  represented  in fig. 3. 

\subsection{Confidence band of the population mean}

  An obvious estimator   of  $\tilde\mu (t)$ is the  smoothed population  mean 
$\overline{S} (t)$.  
The construction of a confidence band for the population mean has  been  recently investigated  \cite{cao, degras}, and we apply to our case  the results  reported  in literature on  the pointwise   confidence band in the normal approximation.    An estimator of the variance of  $\tilde Z(t)$ is
\begin{equation}
\hat \sigma(t) ^2= \frac1{n-1} \sum_{i=1}^n (S_i (t) - \overline{S} (t) )^2
\label{varianza}
\end{equation}
Then the   $1-\alpha$  level confidence band is
\begin{equation}
\overline{S} (t)  \pm \hat\sigma(t)\  z_{1-\alpha/2}\  n^{-1/2}
\label{cof.band}
\end{equation}
where $z_{1-\alpha/2}$ is the standard normal quantile.  

 \subsection{Feature extraction}
In the model eq. \ref{modellonormalapprox1} of the raw data series 
we consider the population mean, defined by 
 \begin{equation}
 \overline Y (t)= \frac1{n} \sum_{i=1}^n \  Y_i(t) 
 \end{equation}
that can be rewritten as 
\begin{equation}
 \overline Y(t) = \tilde\mu(t) + W(t)
 \label{modello2}
 \end{equation}
 where the error term is now
\begin{equation}
W(t)= \frac 1 n \sum_{i=1}^n \ \tilde Z_i(t)
\label{error.term}
\end{equation}
The  estimate of $\tilde\mu (t) $  can now  be considered as a problem of trend extraction from $\overline Y (t)$. 
This is a non smooth function presenting  several maxima and minima, and our aim is to extract the significant ones 
(feature extraction). The same problem has been considered in different applications (see for instance \cite{song}) and it
 consists essentially in constructing  a confidence band for $\tilde\mu (t)$  and for its derivative. 
  The   \verb#R#  package \verb#SiZer#  (Significance of zero crossings of  derivatives) \cite{sizer, chauduri} has been developed for the exploration of structures in curves.
This package uses a locally weighted  polynomial regression  centered at each point  $t_j, \ j=1,...,k$ of a grid, a suitable subset of the time values; a kernel gives a weight to each point; for each grid point the    estimated parameters provide  the  local trend and   slope.  The  weight has  the form 
\begin{equation}
 w_j(t) =  \frac{K( ( t - t_j ) / h )}{ \sum_t  K( ( t - t_j ) / h )}
 \label{kernel}
 \end{equation}
where $K$ is a symmetric  function concentrated near zero, for instance a Gaussian, and $h$ is the bandwidth.  We use  a second order polynomial
\begin{equation}
P_j (t) = \beta_0^{(j)} + \beta_1^{(j)} (t- t_j) +  \beta_2^{(j)} (t-t_j)^2 
\label{polyn}
\end{equation}
and the parameters $\beta_0^{(j)},  \beta_1^{(j)},  \beta_2^{(j)}$ are estimated minimizing  
for any $j$  the quantity
\begin{equation}
 \sum_{t} ( \overline Y(t) - P_j(t) )^2 w_j (t) 
 \label{costo}
 \end{equation}
The value  of $h$ acts as  a  smoothing parameter.  The   coefficients of zeroth order  $\beta_0^{(j)}, j=1,...,k$  are an estimator of the level; similarly the coefficients of first order $\beta_1^{(j)}$ are an estimator of the derivative.

 \begin{figure}[!ht]
\begin{center}
 \vspace*{8pt}
 \includegraphics[width=0.9\textwidth]{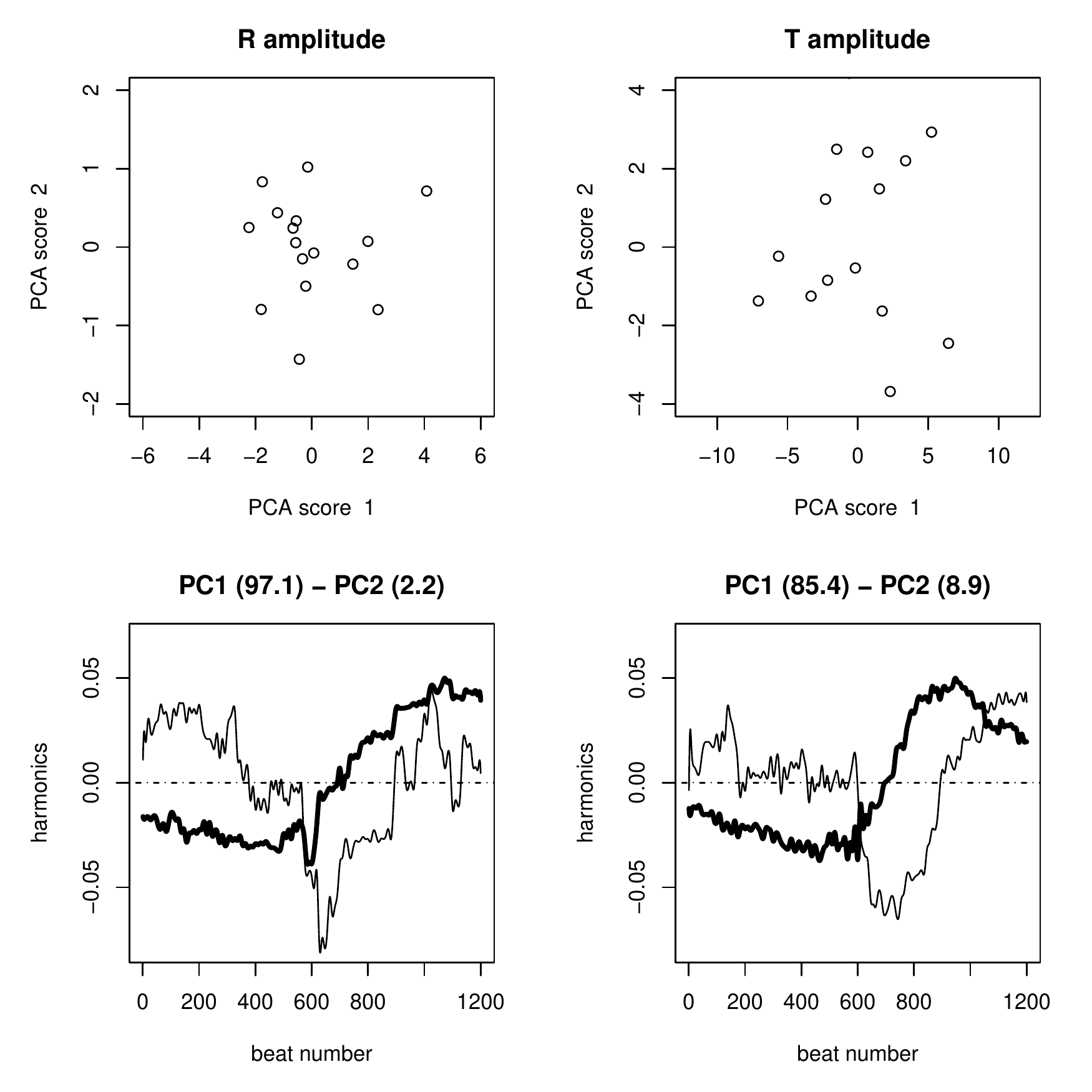}
 \end{center}
\caption{First row: R and T  loadings on the two first principal components (PC1 and PC2); 
second row: normalized  PC1 (thick line) and PC2 of R and T amplitudes  (after mean subtraction).}
\label{fig.pca}
\end{figure}

 \section{Results}
 We  have analyzed    16  normal subjects  (aged 45 $\pm 15$ years) who underwent  the test performed according to the Bruce protocol in a preceding study of our group \cite{curione2008} to which we refer for clinical details.
The RR intervals at rest range from 400 to 800 ms;  the R amplitudes range from 1 to 3  mV and the T amplitudes range from 0.2 to 1 mV.  The series are subjected to data registration and normalization. The smoothing is performed using 
a splines basis   constructed over a grid   of 135 knots. 
The resulting  RR, R and T  series are  plotted in  fig. 2  for  the  16  individuals (color on line).  

The principal component analysis for our dataset is extremely effective, since the explained variance with  the first two components  PC1 and PC2 of  R wave  is 99\%  and  of T wave is 94\%.
The correlation between  $c_{i,1}$  and  $c_{i,2}$  of R and of T are non significantly different from zero  as expected (fig.3, first row). Both  PC1 components  reflect the mean trend  near the acme (beat number 600): a local minimum  for R and a maximum for T. Both PC2 components intersect the zero line at acme, reflecting a contrast effect between exercise and recovery (fig. 3, second row). 

As already known each individual RR series shows a well defined V shaped profile; this profile is non symmetric, with the stress phase (before the acme) and the recovery phase (after the acme) having different slopes.  The R and T individual series on the contrary show several local extrema, most of which should be not significant, reflecting individual  random responses to exercise (fig. 2).  The inspection of the population mean (the  thick line   superimposed  on the plots of R and T in fig. 2), shows the   simultaneous occurrence  of a dip in R and of a bump in T  series  during the early recovery phase. We have conjectured that this evolution, not  previously reported,  characterizes the  normal response to  exercise. In order to assess the significance of these features we  test the hypothesis that the  population means   of  R and T  series  are constant in time. In other words   the  null  hypothesis  is that the data do not contain any information different from noise.  In  fig. 4 we  represent the confidence band with  $\alpha=0.05$ and $z_{1-\alpha/2}=1.96$,   computed using eq. \ref{cof.band} and the constant straight line of height equal to the time mean. In both  R and T  series just after the acme  the   confidence band  does not include the constant  line,  so we conclude that  respectively the  dip  in R data and the bump  in T data  are  significant.

\begin{figure}[!th]
\begin{center}
 \vspace*{8pt}
 \includegraphics[width=0.9\textwidth]{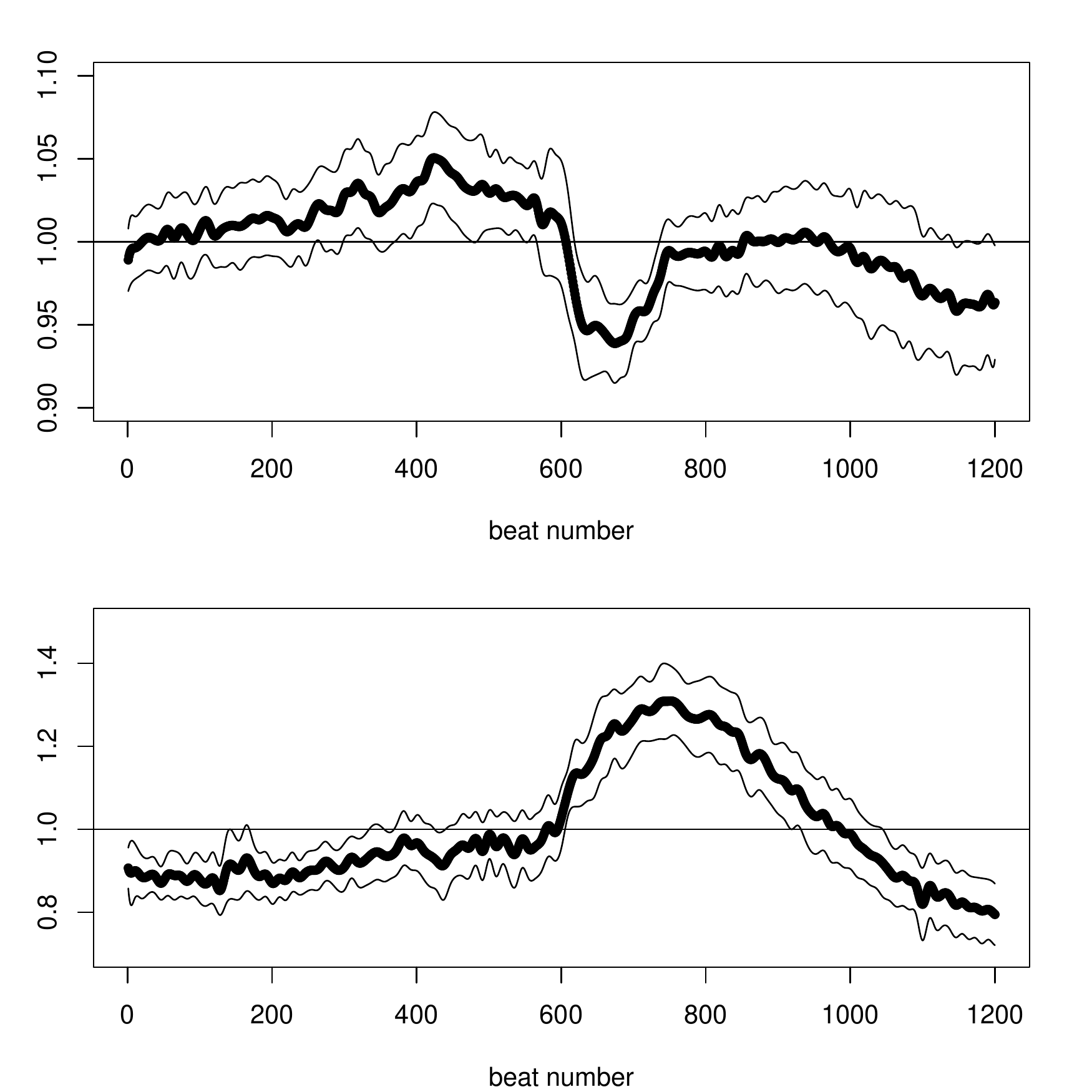}
 \end{center}
\caption{ Confidence   band   of  level $95 \%$  of  the  population mean of R (top) and T (bottom)  amplitude (adimensional units)}
\label{fig.conf}
\end{figure}

\begin{figure}[!ht]
\begin{center}
 \vspace*{8pt}
 \includegraphics[width=0.9\textwidth]{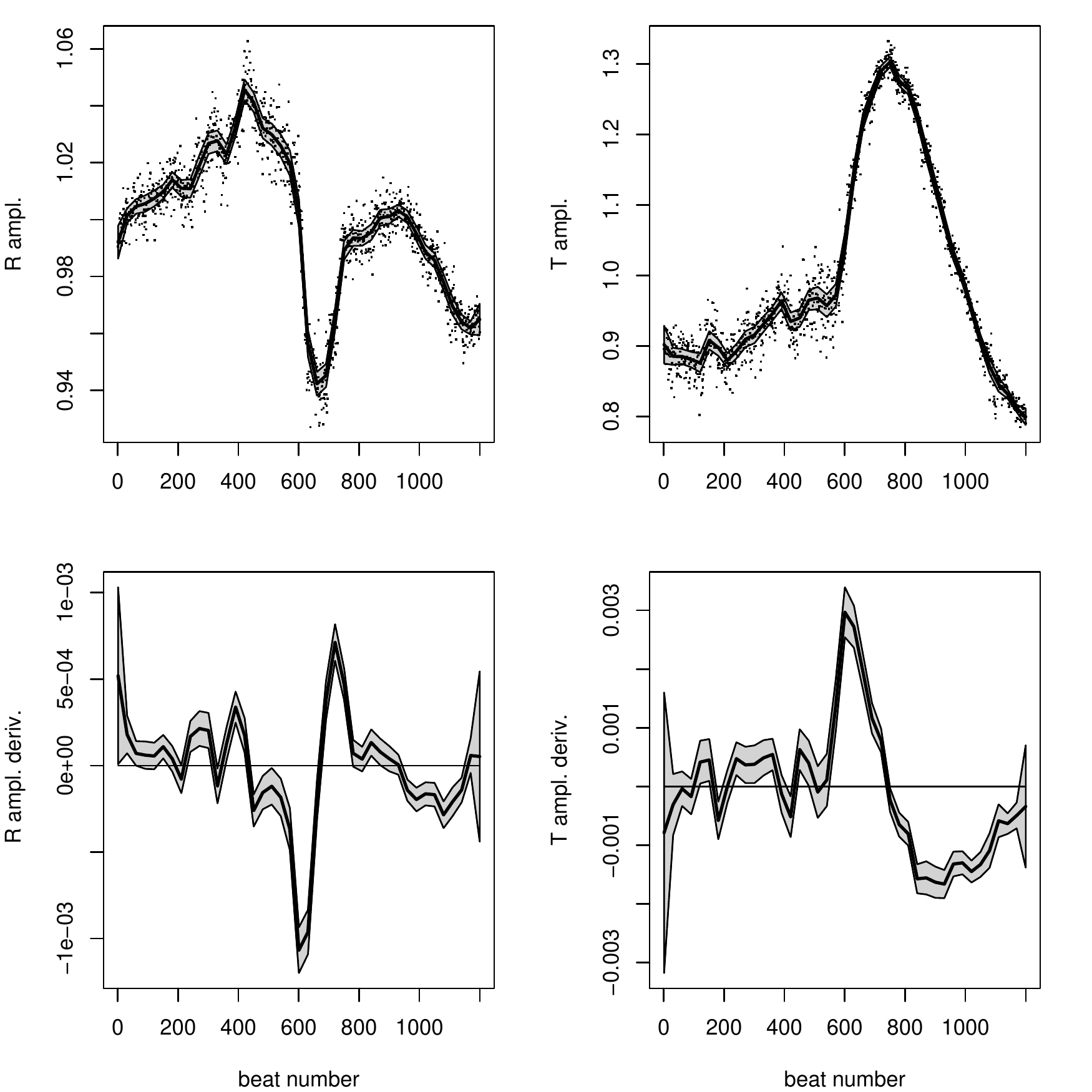}
 \end{center}
\caption{First row: R and T   population mean amplitude  profile (adimensional units)    with their 95 \% confidence band (gray), and  observed  values (dots);
Second row:  estimated derivatives and confidence band. The bandwidth is $h=20$.}
\label{fig.sizer}
\end{figure}

 A second approach is based on the analysis of  the  residual  variability in the  raw population mean, obtained using the eq. \ref{modello2}. 
For the  localization of  a local extremum  in a series we adopt the method based on  the zero line crossing of the derivative: an extremum  is significant if the derivative is different from zero and has  opposite signs before and after the zero crossing.  
If  the  zero level line is contained in the confidence band  of the derivative,  local maxima and minima are not significant.  
The   SiZer package  provides  a confidence band for the population mean $\tilde\mu(t)$  and for its derivative, according to the method described in the previous section. 
As expected the  two main features of the population mean  of R and T series, 
 the dip of R and the bump of T, occurring just after the acme,  are significant (fig. 5).

\section{Discussion}
Two main sources of variability are present in the  exercise ECG data.  The first one is the inter individual variability, which reflects the different individual responses to the stimulation during the test. 
 The second  source of variability  is  due to  the random  fluctuations in time that are present in  each signal, mainly due to the error measurement.  Both these effects produce  the peaks and valleys observed  in each individual series of   R and T amplitudes.  

We have performed  a test on the  trend of  the population mean of  R and T amplitudes   in the framework of functional data.
This test  based on  the confidence band of the population mean  has revealed that the T amplitude    has  a  clear  response  to stress, consisting in a bump after the acme. 
To the best of our knowledge this result is new.  This  test has also  provided evidence of a concomitant  dip  in the R  amplitude, that was described in \cite{curione2008}. The second  test conducted using the feature extraction of the  raw population mean  confirmed these results, revealing that both  the derivatives of R and T amplitude have a significant zero crossing.   

A comparison of the two  normalized population mean profiles  (fig. 4) shows that the  amplitude of the maxima of  the T series   is greater than   the corresponding minimum of the R series. 
These different findings  in R and T behavior  could  reflect a  weaker  response of the R wave to exercise or   different individual timings of the dip in R amplitude with respect to the acme.   
Our dataset is characterized by a small number $n=16$ of individuals, and a large numbers $m=1200$ of observations. Unfortunately theoretical results such as \cite{cao, degras} concern  asymptotic behavior for $m\to\infty,\ n\to\infty$ and finite  sample size approximation results are not available. Our results should be confirmed by a larger sample of individuals. An interesting related  open problem is to provide confidence intervals  of the time location  and of  the amplitude of the extrema, as related to  measurement noise and   inter individual variability.   The normalization of the data series has the effect of reducing the inter individual variability (each series has time mean equal to 1), allowing to focus on  the  time variability. An open problem is to investigate the population mean of non normalized data, for which the large inter individual variability  could mask  some of the  observed effects. 

 The principal component analysis has revealed that 
 a  very large  proportion of variance is explained by the first two   components. This allows a simple bidimensional representation of the individuals, that could be  of  clinical use, for instance in  finding abnormal behavior and in clustering. 
 
Our results  suggest that the population profiles of R and T waves undergo significant  opposite directional changes  just after the acme as a normal  response to exercise. 
They  integrate the previous ones obtained on the R wave\cite{curione2008}, thus confirming  the  relationship between electrocardiographic and hemodynamic variables in normals.  This relationship   was observed in some clinical experimental conditions in which endoventricular volume progressively changes, such as during haemodialysis \cite{curione2013}. 

These findings have an  interesting physiological  interpretation and  possible clinical applications.
In the normal subjects here examined  the mean time profile  of R and T  during increase in heart rate is stable and  this finding could be referred to
normal diastolic ventricular function.  Opposite trends in R and T wave amplitude after the acme seem to show telediastolic and telesystolic volume changes respectively, as they present a specular behavior according to the Frank-Starling law.  The mechanism underlying to this phenomenon is complex and still debated, but the role played by changes in intraventricular volume seems to be the most reliable. Increasing telediastolic volume leads to an increase in electrical resistivity due to higher number of red cells in the left ventricular chamber. This mechanism could play a role both in R and in T wave amplitude changes, representing respectively telediastolic and telesystolic volume changes.    The  R and T amplitudes   profiles  could   have a  clinical application to ischaemic patients  identifying  abnormalities in haemodynamic performance during the exercise test \cite{curione2008}.

The present work has several limitations of various types. 
 The detection of QRS and the measure of  R and T amplitudes are influenced by  the noise in the signal, in particular  by the   electromyographic  noise.   We have not tested  the robustness of our findings   with respect to the addiction of noise to the signal.  The performance of the method should be tested also with respect to  different detection algorithms.
During the exercise the  T wave shape undergoes strong modifications, the main of which is the overlapping of  T wave  offset with  the subsequent P wave onset. This prevents us from considering the  energy area of the T wave that similarly to the one of the QRS complex has been usually adopted.  
Our physiological interpretation of the extrema in R and T trends  should be confirmed by  a direct measure of venous return and ventricular volume during exercise.    The   respiratory-related oscillations of the cardiac axis  and of the electrical impedance have been not considered.  The importance of these phenomena  on the shape of T wave was discussed in \cite{lombardi, porta1998mbec}. However,  these effects  occur on  a scale of  a few seconds, while  the  observed  extrema  occur on a larger time scale.

\section{Acknowledgements}Funding source: MIUR, Italian Ministry of Instruction, University and Research.

\section{Conflict of interest}The Authors declare that there is no conflict of interest.

 \bibliography{cammarota.bib}

 \end{document}